# Context-based spectrum sharing in 5G wireless networks based on Radio Environment Maps

Paweł Kryszkiewicz[1], Adrian Kliks[1], Łukasz Kułacz[1], Hanna Bogucka[1],
Georgios P. Koudouridis[2], and Marcin Dryjański[2]

[1] Chair of Wireless Communications, Poznan University of Technology, 60-965 Poznań, Poland.
[2] Huawei Technologies, SE-164 40 Stockholm, Sweden.

Correspondence should be addressed to Paweł Kryszkiewicz;
pawel.kryszkiewicz@put.poznan.pl

## Abstract

Dynamic spectrum sharing can provide many benefits to wireless networks operators. However, its efficiency requires sophisticated control mechanisms. The more context information is used by it, the higher performance of networks is expected. A facility for collecting this information, processing it and controlling base stations managed by various network operators is a so-called Radio Environment Map (REM) subsystem. This paper proposes REM-based schemes for the allocation of base stations power levels in 4G/5G networks, while considering interference generated to a licensed network. It is assumed that both networks have different profiles of served users, e.g., area of their positions and movement, which opens opportunities for spectrum sharing. The proposed schemes have been evaluated by means of extensive system-level simulations and compared with two widely adopted policy-based spectrum sharing reference schemes. Simulation results show that dynamic schemes utilizing rich context information, outperforms static, policy-based spectrum sharing schemes.

## Introduction

Following the well-known adage saying that "the more you have, the more you want", end-user expectations regarding the offered network capacity are continuously growing. This observation is confirmed by numerous forecasts – they clearly indicate that global mobile traffic will continue to grow in the context of future wireless networks and will reach extremely high levels of peta or even exabytes per month (as indicated by Cisco [1]). It is enough to look at the key performance indicators (KPIs) identified by the 5G Public-Private Partnership (5GPPP) initiative for the Fifth Generation (5G) networks to see that indeed, the requirements defined for future wireless networks in terms of expected throughput are extremely high. It is expected that the 5G network should achieve *1000 times higher mobile data volume per geographical area*[1] comparing to the 4G networks. One way to achieve it is via network densification. Another can be found by looking at the fundamental channel capacity formula proposed by C. Shannon. The observed capacity will grow linearly with the bandwidth, whereas logarithmically with the signal-to-noise ratios. From this perspective, it is worth spending much effort to guarantee reliable access to a possibly wide spectrum band. Numerous spectrum measurement campaigns all over the world have emphasized the problem of high spectrum scarcity, and have resulted in proposing numerous solutions jointly falling into the cognitive radio category [2], [3]. In this approach, the traditional static frequency band and license assignment among various stakeholders is replaced by the dynamic spectrum and license granting solutions [4]-[7]. However, the practical deployment of *pure* cognitive radio concept cannot be realized today, due to many technical obstacles, just to mention the unsatisfactory performance of spectrum sensing algorithms as a vivid example.

---

[1] https://5g-ppp.eu/kpis/
[2] It is assumed that this information is sent explicitly by the outdoor UE. However, one may try to deduce these values based on the already existing reports delivered by the moving outdoor UE to the outdoor eNB (of Operator A), which can be then provided to REM, e.g., long-term CQI reports from the same location.

Recent investigations in the area are focused on the application of highly advanced database-oriented solutions that allow for dynamic spectrum management with the required intelligence and precision. For example, REMs [8]-[12] have been treated as a promising solution for cognitive radio systems, as they bypass many problems occurring due to the abovementioned problem of limited performance of spectrum sensing algorithms. Research conducted all over the world in recent years resulted in the rapid development of this technology, as REMs are now highly envisaged as the real enabler of radio environmental awareness (REA) [12], [13]. Together with advanced (e.g., cooperative and multi-agent) spectrum sensing and monitoring techniques, REMs can be foreseen to play one of the significant roles in the future wireless networks. As REMs are, in principle, databases managed by a dedicated engine (e.g., REM manager), such an approach is in line with the recently popular concept of radio access- network virtualization, where network functionalities are separated from the underlying proprietary appliances [14], [15]. The successful implementation of the virtualization of the wireless networks will rely on orchestrating functions with managed storage, databases and hardware elements.

As we have said before, REMs represent a collection of advanced repositories steered by a dedicated manager, which is also responsible for communication with the "outside world". As various types of data may be stored in the databases, the key role of REMs is to deliver accurate and detailed information on numerous features of the ambient environment. In such a case, the new paradigm of "context-aware communication" can be emphasized, where rich context information is utilized for the optimization of the target utilization function (rate maximization, interference management, traffic steering, load balancing and offloading, etc.). The information about available spectrum at a certain location is stored in dedicated repositories and can be accessed by any interested and allowed player (operator, regulator, policy maker) or user/device (such as mobile terminals, base stations). It may be then intelligently merged and associated with other types of information in order to make an optimized decision for a given set of criteria. Various solutions for such databases have been proposed in the literature [16]-[21]. Some specific discussions on REM implementations have been presented in, e.g., [22] and [23]. While the former one focuses on the application of REMs as a tool supporting LTE Railway systems, the latter concentrates on various security aspects. A highly interesting approach to the application of multi-dimensional maps has been proposed in [24]. Moreover, REMs are also considered as a practical tool for the improvement of vehicle-to-vehicle (V2V) communications [25][26].

One of the key aspects associated with the use of such advanced databases is the need for their periodic, continuous and accurate updates. Such a modification of the current status can be made based on continuous channel measurements carried out by, for example, mobile devices or dedicated sensors. Finally, assuming that databases are filled with rich content data, the issue of efficient database access is also essential in order to maintain a high level of protection of the licensed systems, and to provide high-quality service to secondary users.

In this paper, we deal with the application of REMs for efficient spectrum sharing. A 5G-oriented scenario is considered, where one mobile network operator (MNO) owns some spectrum resources in the 3.5 GHz band (although the extension to any other band is straightforward) and offers its services only to outdoor users (i.e., it is not interested in offering dedicated indoor coverage). From this viewpoint, this MNO would like to share its resources with other players, as the limited service area gives to it an opportunity to share the licensed spectrum, and to provide some radio services to indoor users. In this work, we concentrate on proposing new approaches to spectrum sharing in the considered scenario, ranging from relatively static but REM-based licensed network protection to dynamic protection based on detailed context information, i.e., interference reports provided by each licensed transceiver. An interference report provides the measure of how strong the intra-network interference degrades each transceiver transmission. A prospective REM database is foreseen as an entity facilitating the utilization of rich context information in this environment. Each scenario is evaluated using a system-level simulator of 4G/5G networks. The proposed REM-based solutions are compared with two state-of-art regulatory-based solutions, i.e., the Citizens Broadband Radio Service (CBRS) standard [27], and a modification of the Licensed Shared Access standard [28]. Many scientific papers, e.g., [29], cannot be fairly compared with the proposed solutions. In these works full system knowledge and a simplified system model is typically considered. Here below, we use a more realistic model. We identify a promising use-case for spectrum sharing, where the advanced database system can be applied. Moreover, we present five distinct solutions to the identified spectrum sharing problem and test them by means of computer simulations. The influence of delay and accuracy of context information on the networks performance is evaluated. Additionally, the REM–based subsystem architecture is proposed to serve the considered schemes. These constitute the novelty of our contribution.

The paper is organized as follows. First, the considered system model is presented in detail, and the research problem of advanced spectrum sharing in the 5G context with the use of dedicated REMs is formulated. In the following section, five autonomous solutions of the identified problem are proposed and analyzed. Finally, simulation results are provided and conclusions are derived.

## System model and problem formulation

*System model*
Reliable spectrum sharing among interested operators for more efficient resource utilization is one of many significant issues widely discussed and investigated in the last decade in the context of future wireless networks. In consequence, the design of advanced spectrum management solutions is nowadays a subject of intensive research [30]-[32]. Various technical enablers have been proposed, e.g., in [33] – [35], in order to facilitate advanced solutions for spectrum management. In particular, two schemes are worth mentioning: Licensed Shared Access (LSA) and Citizen Broadband Radio System (CBRS) with Spectrum Access System (SAS) [36] – [42]. These two approaches fully rely on the presence of a dedicated spectrum management system utilizing dedicated databases that store various types of context information, used in the entire management process. In case of LSA, the spectrum is shared, based on individual agreements between the operators, while in case of CBRS spectrum is shared by means of a coordinating spectrum access system function. Application of these schemes in the considered scenario will be presented in the next section.

Let us consider the prospective scenario for a 5G network (or beyond), where one network operator, who owns the spectrum, wants to increase its revenue from it by granting payable access to its licensed frequency band. In particular, let us discuss the situation depicted in Figure 1, where two network operators coexist in a certain geographical area. One may identify the so-called outdoor operator (denoted hereafter as Operator A), who is interested in delivering wireless services outdoors only, and who owns the license for the considered spectrum. Its base stations, denoted as OP_A_BS_1 and OP_A_BS_2, are marked with grey dots, whereas one representative User Equipment (UE), OP_A_UE_1 is depicted as a white dot. Note that in our investigation we consider the spectrum around 3.5 GHz, as we concentrate on the 5G scenario to make the analysis more concrete and simulation results more applicable. However, the technical discussion presented in this paper is completely independent of the considered frequency range. Beside the outdoor network, one may identify an indoor network ruled by Operator B, who is interested in delivering access to wireless services to users located inside a building (e.g., office or mall). Its base stations, denoted OP_B_BS_$i$, $i$=1,2,3, are marked with black dots in Figure 1.

One may observe that the nature of services of both operators is completely different. As the indoor network is deployed to serve the traffic of indoor users associated with Operator B, the users of Operator A are outdoors, unable to utilize licensed spectrum indoors. Here, we consider an interesting scenario of Operator A users being public transportation vehicles. Thus, both networks may be close to each other, but still be separated geographically, as high wall-attenuation decreases interference coupling between indoor and outdoor networks. In such a prospective scenario, both operators may benefit from sharing the spectrum originally assigned to Operator A. This is graphically presented in the top-view scheme showed in Figure 1 by splitting of the area into two sections: above and below the dashed line. Operator A is interested only in serving the users located in the bottom area (e.g., in the street). It is not that interested to serve the users outside this region, and is keen to share its spectrum with Operator B, who is offering indoor services (in the area above the dashed line). Let us just mention that there are two specific points marked in this figure (Point A – within the coverage area of Operator A and B– outside of the coverage area of Operator A) that will be used later in the following section.

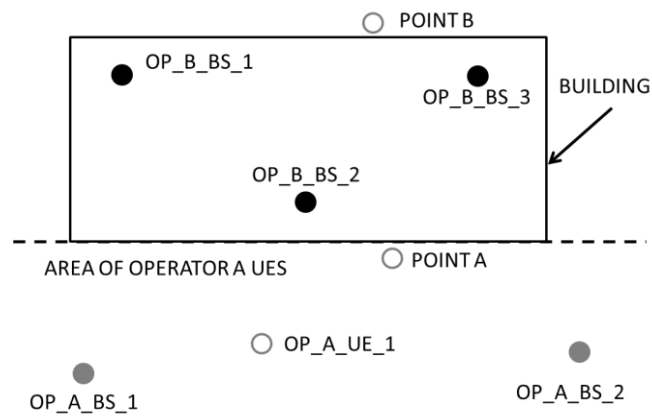

Figure 1. Example of networks deployment: grey points- BSs of Operator A, black points- BSs of Operator B, white points- UE or locations utilized in next sections. Top View.

The issue of mutual interference between collocated networks may need to be controlled in order to guarantee the protection of the licensed users of Operator A. Although wall-attenuation may reduce the level of interference, it is obvious that any new additional transmission in the licensed band (in this case: transmission in the network of Operator B) introduces some interference to the licensed system, i.e., of Operator A, so that the total observed level of distortions increases as well. The amount of interference that may be induced by users of Operator B network can be precisely identified in a mutual agreement between the operators. However, from the technical perspective such an agreement means that the amount of interference can be monitored and controlled by means of some system mechanisms. Furthermore, the lower the allowed interference power to be induced to the outdoor network, the lower the transmission power in Operator B network, resulting in lower available throughput and lower profitability for Operator B. Thus, it is profitable for both operators to define the operation point in an intelligent way, and to dynamically adjust the system parameters to achieve this goal. For example, Operator A could agree to observe some interference power from Operator B, accepting slight, yet unnoticeable by its users, degradation of the total throughput. But again, the practical application of such an approach is feasible only in the situation where advanced controlling systems are deployed, which is described in the following subsection.

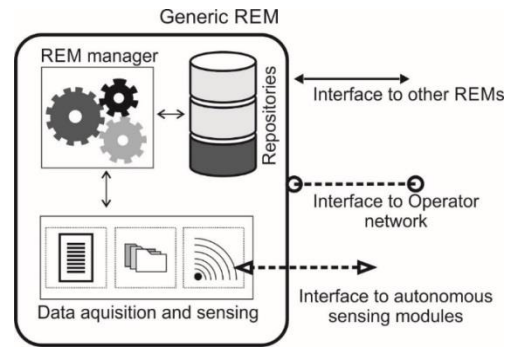

Figure 2. Main functional blocks of an REM-based control system

*REM-based subsystem*

According to our approach of dynamic spectrum management, an REM-based subsystem may consist of three main functional blocks [13] and a set of interfaces, as depicted in Figure 2:

- An REM manager, being the brain or cognitive engine of the entire system, designed for the database management (accessing data, modification of database entries) and data processing; moreover, it performs the whole reasoning with decision making, conformance verification, coordination and control;
- Second, a set of advanced databases (repositories), which may contain various types of context-information and data; it is worth mentioning here that two types of databases may be identified: private and public; in the former case, the database contains some sort of sensitive data for a certain operator (e.g., about its users), and this type of data cannot be shared with other operators; on the other hand, public databases contain information that can be shared among cooperating operators for better network management;
- Finally, the data acquisition and gathering function – the system under the control of REMs may be monitored in various ways; one solution is to deploy dedicated sensing/monitoring modules which deliver necessary data to the REM-manager.
- Beside the functional blocks, a set of interfaces is necessary to guarantee the proper operation of the REM-based subsystem; allowing a high level of abstraction, one may define three types of interfaces: an interface used for message exchange between separate REM-based subsystems, an interface to communicate with the legacy network of the certain operator, and finally a prospective interface for sensing modules for better measurement collection.

In the considered case, we assume that each operator is in possession of its own REM-based spectrum management system, containing a set of private and shared (public) databases. In the latter case, information originated in one network (associated with one operator) may be shared, if needed or at request, with other REM-based spectrum management systems, as in the mutual agreement. In our use-case, a shared REM-based management system (denoted in the figure as *3rd party/OpB REM*) is used for spectrum management inside the building and may be managed by Operator B or theoretically may even be in possession of the building owner (however, the discussion on the benefits and drawbacks of these two solutions is out of scope of this paper). Clearly, each REM-based spectrum management system is only a part of the whole network management system, thus dedicated interfaces between the REM and Network domain have to be maintained.

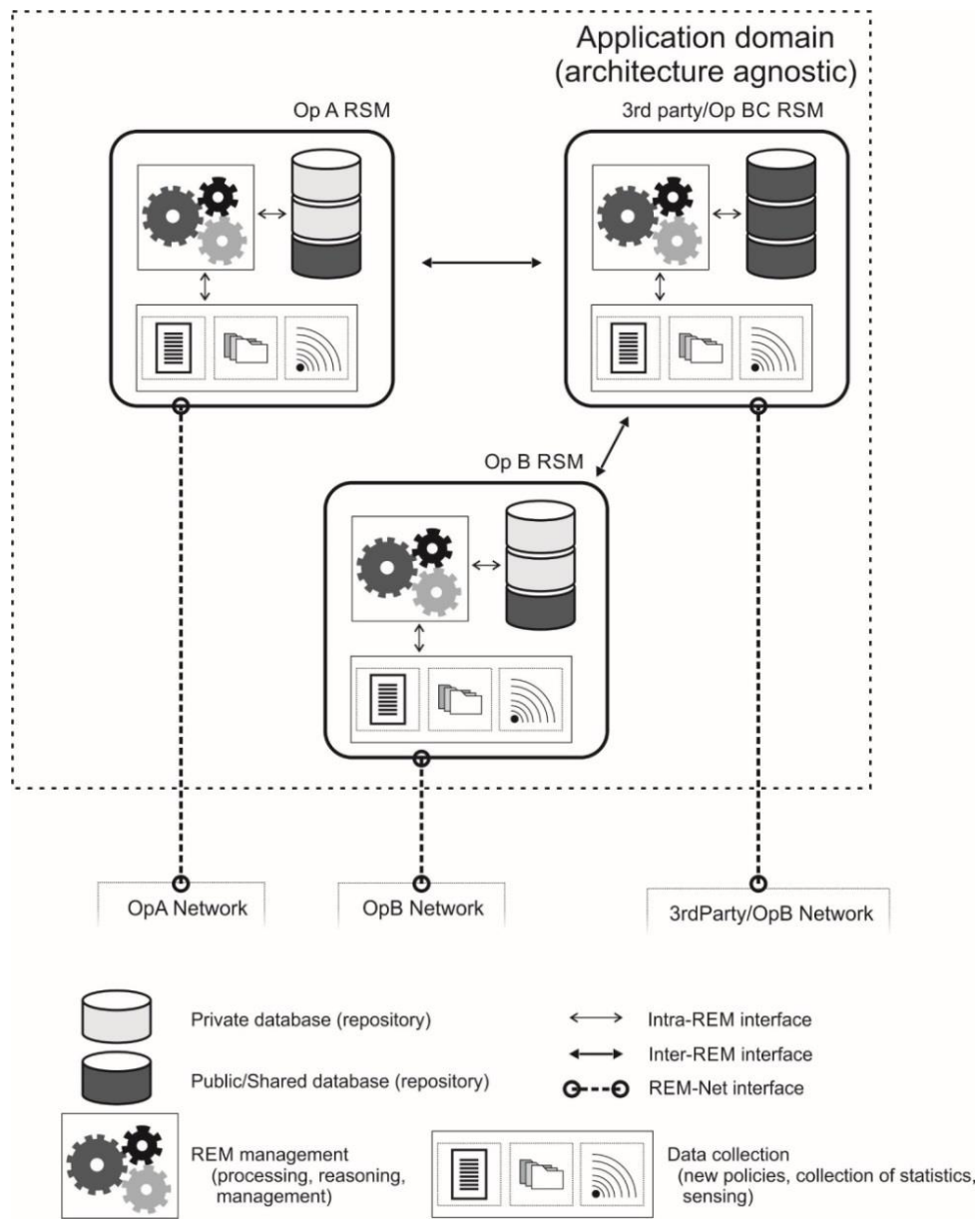

Figure 3. Proposed structure of the database system

*Problem formulation*

In order to formulate the optimization problem, let us return to Figure 1. We consider two coexisting networks (indoor and outdoor), and it is assumed that all base stations have fixed locations and use the same frequency channel. Our goal is to allow advanced spectrum sharing with the purpose of increasing the income of Operator A (by granting dynamic access to its licensed spectrum) and guaranteeing to Operator B the reliable deployment of its indoor network. It is assumed that Operator A accepts some performance degradation in its network, but still expects that its users are served in a fair way. We assume that the only degree of freedom in the maximization of indoor rate, while maintaining Operator A users' Quality of Service at an agreed level is in the indoor base stations power allocation. Time domain-based resource distribution is not considered, as it would require centralized scheduling, not practical in contemporary network design. Additionally, as high traffic asymmetry is observed in cellular networks, it is assumed that spectrum sharing is employed in the downlink (DL) only. The protection of Operator A UEs, in the area served by it, can utilize various amounts of context information. It can vary from parameters describing propagation conditions and operation region, to precise interference reports provided by Operator A. The collection of such varying information, its processing and control of the power of indoor BSs are tasks that can be carried out by the REM.

## Considered solutions for advanced spectrum sharing in 5G networks with the utilization of REMs

In this section we consider five solutions for advanced spectrum sharing. While two of them are based on existing spectrum sharing standards, the others are proposed in this paper and are using a REM-subsystem. These five solutions vary in terms of their dynamicity – we start from a fully static scenario and finish with the fully dynamic approach. At the end of each sub-section, we concisely discuss the role that REM-subsystem may play in such a case.

A) Modified LSA: static, regulatory-based protection

In this scenario, we assume that the allowed coverage area of Operator B is defined and known. In the considered example, it is the whole area of a building. This is an improvement over the standard LSA scheme. All other locations, i.e., points outside the building, have to be protected from interference. As such, we can define a belt around the building, in which any potential user of Operator A has to be fully protected. In other words, there is no adaptation in time between indoor BSs of the transmit parameters. The maximum values of the transmit powers (Effective Isotropic Radiated Powers - EIRPs) are calculated once to protect the whole area outside the building, no matter what is the current location of the outdoor UEs of Operator A or the wanted signal received power. As such, the protection will be defined with respect to the thermal noise floor similarly as in [28].

The maximum transmit power of each base station is calculated *a priori* to transmission. First, the protection belt is defined as a set of points $(x_n, y_n)$ for *n*=1,…,*N* located just outside building walls. The transmit power for each indoor BS, α, is calculated independently. We assume that the outdoor UEs may receive an interfering signal whose power is similar to the thermal noise power (defined as $\sigma_N=kTB$, where *k* is the Boltzmann constant, *T* is the ambient temperature and *B* is the operational bandwidth) increased by the protection belt margin *Γ*. Protection belt margin Γ can be interpreted as the ratio of thermal noise power (assuming 0 dB noise figure) to the maximum interference power that can be inducted at location $(x_n, y_n)$ by a single indoor BS. Theoretically, $\Gamma$ in logarithmic scale can take any possible real value, however, in practice, it will vary from, e.g., 0 to -60 dB.

Thus, the maximum interference power (expressed in dBm) received by an outdoor UE from a given indoor BS, which is still acceptable by Operator A, is defined as:

$$P_\alpha^{RX}(x_n, y_n) = \sigma_N - \Gamma = -174 + 10\log_{10} B - \Gamma, \qquad (1)$$

where the value -174 dBm refers to the thermal noise power at 1 Hz at typical room temperature (300K). In order to calculate the maximum transmit power for the indoor base station, we need to consider at least the classical propagation relations, i.e., the Friis transmission formula which takes antenna gains and path loss (PL) into account. Thus, the indoor BS maximum transmit power in dBm is calculated as

$$P_\alpha^{TX}(x_n, y_n) = -174 + 10\log_{10} B - \Gamma - G + PL, \qquad (2)$$

where $G$ is BS TX antenna gain in dBi and PL is pathloss from BS $\alpha$ to location $(x_n, y_n)$ in dB. The BS TX power is constrained by this formula at each point of the protection belt. A proper choice of protection belt margin is required in order to provide sufficient protection of Operator A UEs. According to [28], the required interference $I$ to noise $N$ power ratio is -6 dB, i.e.,

$$(I)_{dBm} - (N)_{dBm} = -6\ dB. \qquad (3)$$

The noise power is understood there as thermal noise $\sigma_N$ increased by noise figure (e.g. 9 dB). The interference is $P_\alpha^{RX}(x_n, y_n)$ increased by UE receiver antenna gain (typically 0 dBi). As such, the above equation can be rewritten as

$$\{P_\alpha^{RX}(x_n, y_n) + 0\ dBi\} - \{\sigma_N + 9\ dB\} = -6\ dB \qquad (4)$$

giving after simplification

$$P_\alpha^{RX}(x_n, y_n) = \sigma_N + 3. \qquad (5)$$

When comparing (5) with (1), one can see that LSA-based protection using [28] sets $\Gamma$ to -3 dB. An additional constraint is the maximum transmit power supported by the device or limited by legal conditions $P^{MAX}$. Finally, the allowed power of BS $\alpha$ is the minimum of the values calculated at each point of the protection belt, i.e.,

$$P_\alpha^{TX} = \min\ \left(P^{MAX}, P_\alpha^{TX}(x_1, y_1), \dots, P_\alpha^{TX}(x_N, y_N)\right). \qquad (6)$$

*REM subsystem role*

This scenario can be treated as a reference one, utilizing a regulatory-based, modified LSA scheme. The power allocation is calculated before indoor network deployment. No REM is needed once the protection level is fixed. Moreover, no knowledge on the outdoor BSs or UEs is needed. However, such an approach requires the interference to be limited even in points not used by the outdoor network, e.g., Point B in Figure 1. Additionally, $\Gamma$ equal to -3 dB assumes the worst-case scenario of a noise-limited outdoor network. As such, the calculated power of indoor BSs is expected to be relatively low. However, REM may be used if one could allow for changes of the protection setup.

B) CBRS-based

This scenario is based on the CBRS standard [27] assuming that the outdoor network represents the second (protected) tier of users, i.e., Priority Access License (PAL) and the indoor network belongs to the third tier of users, i.e., General Authorized Access (GAA). In this standard, similarly as in our setup, the GAAs should protect PAL devices from interference while accepting the whole incoming interference. The transmission of PAL devices is protected in the so-called "PAL protection area" being all locations where the received power of the wanted signal exceeds -96 dBm/ 10 MHz. At each of these points the cumulative interference caused by GAA devices should not exceed -80 dBm/10 MHz.

*REM subsystem role*
The REM subsystem is not used in this scenario. The Spectrum Access System in CBRS uses information on device location and a propagation model to establish a PAL protection area and allowed GAA transmit power. Although it can be accomplished using information stored in a database, no context information regarding indoor network QoS or differences between both networks operation conditions is used.

C) Semi-static, REM-based protection

This scenario is a slight modification of the first one, where regulatory-based protection belt margin $\Gamma$ is used. In this scenario, the protection belt margin is optimized according to interference conditions of the outdoor network. The proposed solution is to use REM to collect and analyze the throughput of users in the area where spectrum sharing is implemented. The 10$^{th}$ percentile of users' throughputs (averaged for each user over some observation window) is compared when the indoor network is on and off. The protection belt margin is optimized in order to provide reduction of these users' rate by, e.g., 10%. This should allow Operator A to keep the high quality of service even for users located close to the building, while allowing the indoor BSs to transmit with higher power (with respect to the "Modified LSA" scenario).

*REM subsystem role*
REM utilizes context information of the outdoor users' rate connected with specific locations. It influences all BSs power by the same factor, i.e., $\Gamma$. The disadvantage of this approach is slow adaptation to the changing conditions of outdoor users. It is possible that at some time instant, an outdoor UE will approach the building walls. The mean user throughput stored in REM will change slowly (depending on the averaging window utilized), causing a temporal degradation of the outdoor UE rate. The protection is based on some statistics, unlike dynamic protection presented in the last approach.

D) Semi-static, REM-based protection with known protection area

This scenario can be seen as an extension of the previous one with additional context information. In this case, we only want to protect a selected area (not the entire belt around the building). The rationale of this approach is to show that we can benefit from the knowledge on the activity area of Operator A users. This can be based on the nature of services provided by Operator A, e.g., for communications with public transportation vehicles the UEs will appear only on the road and at stops. In the example shown in Figure 1, the area of possible outdoor UEs appearance is below the dashed line. Although Point A has to be protected, similarly as in the previous scenario, Point B does not belong to the protection area and the interference is not limited there.

The indoor BS power is calculated according to the same formulas as in the "Modified LSA" scenario. The only difference is the set of points in the protected area $(x_n, y_n)$. Additionally, the protection belt margin $\Gamma$ in this scenario can be different than in the previous scenario, although it is established in a similar way.

*REM subsystem role*
The REM database is used here not only to collect and process outdoor user rates. In addition, it can collect information on the past locations of outdoor users and based on it establish the protection area. It should be passed to the function calculating the transmission power of indoor BSs according to (6).

E) Dynamic, REM-based protection

In the final scenario, we consider the application of the outdoor BSs power optimization algorithm based on the interference reports passed to the REM databases. It is assumed that the outdoor UEs, in parallel to Channel Quality Indicator (CQI) reporting, send information on the ways the out-of-system (in our case: indoor BSs-originated) interference could be increased or should be decreased[2]. Let us assume that a given outdoor UE n, located at position $(x_n, y_n)$ has measured the wanted signal power **S** (vector with a single power value for a single subcarrier), noise power $\sigma_N^2$, intra-network interference power $\mathbf{I_{IN}}$ (i.e., originating from other outdoor transmitters within the Operator A system), and inter-network interference $\mathbf{I_{OUT}}$ (i.e., originating from indoor transmitters of Operator B and observed by the outdoor UEs of Operator A). The signal to interference and noise ratio (SINR) is $\frac{S}{\sigma_N^2+I_{IN}+I_{OUT}}$. It is assumed the UE can estimate its internal noise separately. The separation of intra-network and inter-network interference can be based, e.g., on the measurement of the neighboring cells' synchronization signals (orthogonal, Zadoff-Chu sequences). In the simulator, full knowledge of incoming signals is assumed. Let us denote *R*(SINR) as a function mapping the vector of SINR values to the rate achievable in this case. There are many different approaches to modeling this function, e.g., based on the Shannon formula, narrowband CQI (sum of rates achievable in a single resource block), wideband CQI (single CQI for the whole bandwidth), etc.[43]. In the simulator, the approach based on narrowband CQI is used. The practical achievable rate depends on the scheduler's operation in BS (i.e., which RBs and Modulation Coding Scheme (MCS) is chosen).

Obviously, any non-zero $\mathbf{I_{OUT}}$ would cause some degradation of the achievable rate. Operator A has to agree that for the worst-case user, $\Psi$ percent of the maximum rate (for $\mathbf{I_{OUT}}$=0) is only achievable. The other limitation (not considered here) is the possible degradation of SINR by, e.g., 1 dB. The UE has to find out interference multiplier $\beta_n$, such that

$$\frac{\Psi}{100} R\left(\frac{S}{\sigma_N^2+I_{IN}}\right) = R\left(\frac{S}{\sigma_N^2+I_{IN}+\beta_n I_{OUT}}\right). \tag{7}$$

Observe that only the UE can find the $\beta_n$ value. It has access to all incoming signals. BSs or REM only have access to the reported measurements, e.g., CQI values. As mentioned previously in a footnote, the utilization of these measurements for the estimation of interference reports is possible, but not studied here. Note that β$_n$ indicates how much the inter-network interference should be changed (indoor BS power decreased or increased) for the outdoor UE to achieve Ψ percent of its maximum rate, e.g., 90. Observe that $\beta \epsilon (0, \infty)$. Values below 1 mean the interference has to be decreased. Values higher than 1 mean that the interference can be (but not always has to be) increased. In the simulator, this equation is solved iteratively.

---
[2] It is assumed that this information is sent explicitly by the outdoor UE. However, one may try to deduce these values based on the already existing reports delivered by the moving outdoor UE to the outdoor eNB (of Operator A), which can be then provided to REM, e.g., long-term CQI reports from the same location.

The value $\beta_n$ in dB scale is denoted as $\beta_n^{dB}$. BS passes it to REM along with user location $(x_n, y_n)$. This may introduce some additional processing/transmission delay (in the simulation, the minimum value of 1 ms is assumed).

The collected "interference reports" can be used by the indoor REM to adjust indoor BSs power. Obviously, the decision on the indoor BS transmit powers is taken on the basis of outdated data (caused by the delay of $\beta_n^{dB}$ reporting from outdoor UE to outdoor BS and inter- REM communication). This operation has to be performed cyclically as the interference reports of outdoor UEs change over time. The indoor REM contains a collection of reports: $(x_1, y_1, \beta_1^{dB}), \ldots, (x_N, y_N, \beta_N^{dB})$ for N outdoor UEs. It does not have any precise channel measurements to outdoor UEs. However, based on an assumed pathloss model, the interference power (in mW) from all indoor BSs, i.e., $\alpha \in \{1, \ldots, \alpha_{IN}\}$, observed at location $(x_n, y_n)$ of $n$-th outdoor UE is estimated as

$$I_n = \sum_{\alpha=1}^{\alpha_{IN}} 10^{\frac{-PL_{n,\alpha} + G_n + G_\alpha}{10}} P_\alpha^{TX}, \tag{8}$$

where $PL_{n,\alpha}$ denotes the estimated pathloss between BS α and location $(x_n, y_n)$ in dB, $G_n$ is RX antenna gain of victim UE, $G_\alpha$ is TX antenna gain of BS α and $P_\alpha^{TX}$ is the currently allocated power of this BS in mW. Constructing N-by-$\alpha_{IN}$ matrix **W** with entries

$$W_{n,\alpha} = 10^{\frac{-PL_{n,\alpha} + G_n + G_\alpha}{10}} \tag{9}$$

it is obtained that $\boldsymbol{I} = \boldsymbol{W} \boldsymbol{P}^{TX}$ where $\boldsymbol{I}$ is an N-long vector of the estimated interference power, and $\boldsymbol{P}^{TX}$ is an $\alpha_{IN}$-long vector of the current indoor BS power. The vector of the maximum allowed interference power $\tilde{\boldsymbol{I}}$ is calculated using interference reports as $\tilde{I}_n = 10^{\frac{\beta_n^{dB}}{10}} I_n$. Obviously, many different combinations of $P_\alpha^{TX}$ cause the same estimated RX interference power. The optimization of the new power allocation vector $\widetilde{\boldsymbol{P}^{TX}}$ for indoor BS power is defined as

$$\max_{\widetilde{P_\alpha^{TX}}} f(\widetilde{P_\alpha^{TX}}) \tag{10}$$

$$\text{s.t. } \boldsymbol{W} \widetilde{\boldsymbol{P}^{TX}} \leq \tilde{\boldsymbol{I}}$$
$$0 \leq \widetilde{\boldsymbol{P}^{TX}} \leq P^{MAX}.$$

The goal function $f(\widetilde{P_\alpha^{TX}})$ can be defined in many ways. Our basic approach is

$$f(\widetilde{P_\alpha^{TX}}) = \sum_{\alpha=1}^{\alpha_{IN}} \widetilde{P_\alpha^{TX}}. \tag{11}$$

This is a linear programing problem that maximizes the total indoor BS power, while keeping it below the device specific/legal limit $P^{MAX}$.

Another possible goal function is

$$f\left(\widetilde{P_\alpha^{TX}}\right) = \min_{\alpha=1,\dots,\alpha_{IN}} \widetilde{P_\alpha^{TX}}. \tag{12}$$

It causes the maximization of the minimum indoor BS power.
The third tested option is the maximization of sum of logarithms of TX powers, i.e.,

$$f\left(\widetilde{P_\alpha^{TX}}\right) = \sum_{\alpha=1}^{\alpha_{IN}} \log\left(1 + \widetilde{P_\alpha^{TX}}\right). \tag{13}$$

While function (11) usually maintains some BSs turned off (i.e., 0 mW allocated power), function (12) maintains all BSs on. The third function, i.e., (13), finds balance between both of the above mentioned ones. It is solved by observing that the sum of the log function is equal to the logarithm of product. As the logarithm is a monotonically increasing function, the goal function can be defined as

$$f\left(\widetilde{P_\alpha^{TX}}\right) = \sqrt[\alpha_{IN}]{\prod_{\alpha=1}^{\alpha_{IN}}\left(1 + \widetilde{P_\alpha^{TX}}\right)} \tag{14}$$

being a convex optimization problem [44].
Most importantly, the relative nature of the interference report constitutes a great advantage of this approach. Even if there is some error in the pathloss estimation, i.e., erroneous entries in matrix **W**, causing error in interference vector **I**, the generated interference will iteratively converge to the correct values.

*REM subsystem role*
This approach requires a relatively high amount of context information to be collected and processed by REM. Each outdoor UE periodically sends interference reports. The low delay, high periodicity and high accuracy of interference reports can potentially improve Operator A UEs protection. However, this places high requirements on the amount and latency of control information passing. The advantage of this method over the two previous REM-based approaches lies in the dynamic adjustment to the interference situation of outdoor UEs. In the case of semi-static REM-based approaches, the protection of outdoor UEs is statistical, based on long-term measurements.

## Simulation results

In order to compare the five presented scenarios (two regulatory-based and three proposed, REM-based solutions), and to show the benefits of using REM and its embedded intelligence, extensive system level simulations of the 4G/5G network have been carried out. The example test environment spans the area of 100x130 m where L-shaped building is located as shown in Figure 4. There are 7 base stations: two outdoor BSs deployed by Operator A (numbered 1 and 2), and five BSs located indoors (numbered 3,…,7). The model of the 1-floor building is based on the floor plan of the building of Faculty of Electronics and Telecommunications, Poznan University of Technology. All cells are operating at the carrier frequency of 3.5 GHz in the same channel of 20 MHz bandwidth. Only the downlink of the FDD LTE-A/5G system is considered. It is foreseen that 5G networks will be partially backward compatible with existing LTE-A networks, e.g., as a result of OFDM-modulation with similar numerology. Therefore, it is assumed that on average 50% of indoor UEs will use 5G technology. The other UEs will use LTE-A technology. Each UE reports one of 15 CQI values per each RB separately. SISO transmission is assumed. The SINR values calculated per each subcarrier are mapped on to the proper CQI values using Exponential

Effective SINR Mapping (EESM) as in [45][46]. At the time of simulations, the exact specifications of 5G modulation and coding has been unknown. As such, the general assumptions of increased spectrum

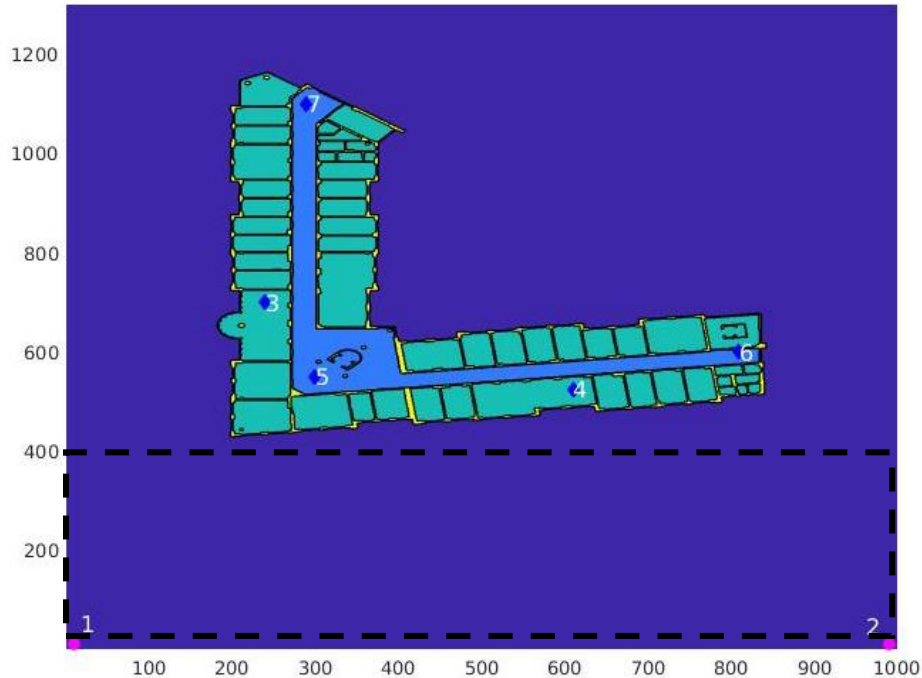
Figure 4. Location (fixed) and numbering of base stations

utilization and efficiency has been mimicked by an addition of 8 extra RBs to 100 utilized in standard LTE-A, and 5G UEs rate increased by 5% (in comparison to 4G UE having the same SINR) as a result of improved coding. Both RX and TX antennas are assumed to have 0 dBi gain. UEs' noise figure is of 9 dB. Each BS can transmit with the maximum power of 21 dBm, which is typically reduced by a chosen scenario for the indoor BSs. Inter-Cell Interference Coordination (ICIC) has been applied using the Soft Frequency reuse scheme. The available bandwidth is divided into 3 parts. In a given cell, one of these parts has 4 times increased power as in [47]. The power in the whole band is normalized to achieve the required total power of, e.g., 21 dBm. While outdoor BSs are located 10 m above ground level, indoor BSs are located below the ceiling, at 3 m. If not stated differently, there are 25 indoor users equally distributed inside, with 80% of them static (speed for Rayleigh channel: 0.36 km/h) and 20% walking at the speed of 3 km/h. Additionally, there is a cluster of 10 static indoor users located close to BS 3, resembling a meeting room scenario. As for the outdoor network, there are 15 users distributed equally in the dashed rectangle from Figure 4. The propagation channel follows the Winner II model [49] with the 7-paths Extended Pedestrian A model. Each tap follows the Jakes model.

At the beginning of each simulation run each randomly located UE is assigned to one BS. The one with the highest power of incoming signal is used. Afterwards, 1 second of transmission is simulated, with CQI reported by UEs every 1 ms and scheduling performed every 1 ms. The scheduler uses a proportional fair algorithm [48] with exponential moving average of the past UE rate using a smoothing parameter of 0.5. If not stated differently, 200 independent iterations have been performed. In each iteration 1000 ms time horizon has been simulated, over which pathloss and Rayleigh channel coefficients change continuously according to the chosen UE speed.

The solution of the optimization problem in dynamic, REM-based protection depends on the goal function. While (11) is solved by the *linprog* function in Matlab, (12) requires the *minimax* function from the same software. In the case of (13) a CVX toolbox [44] is used.

First, in order to allow a fair comparison of different scenarios, the protection belt margin Γ used in "Semi-static, REM-based protection" has to be fixed. It is done by running simulations for Γ={-50, -40, -30, -20, -10, -3, 0} and finding outdoor UEs $10^{th}$ percentile rate that is decreased by about 10 % in comparison with the case of no indoor BS-originating interference.

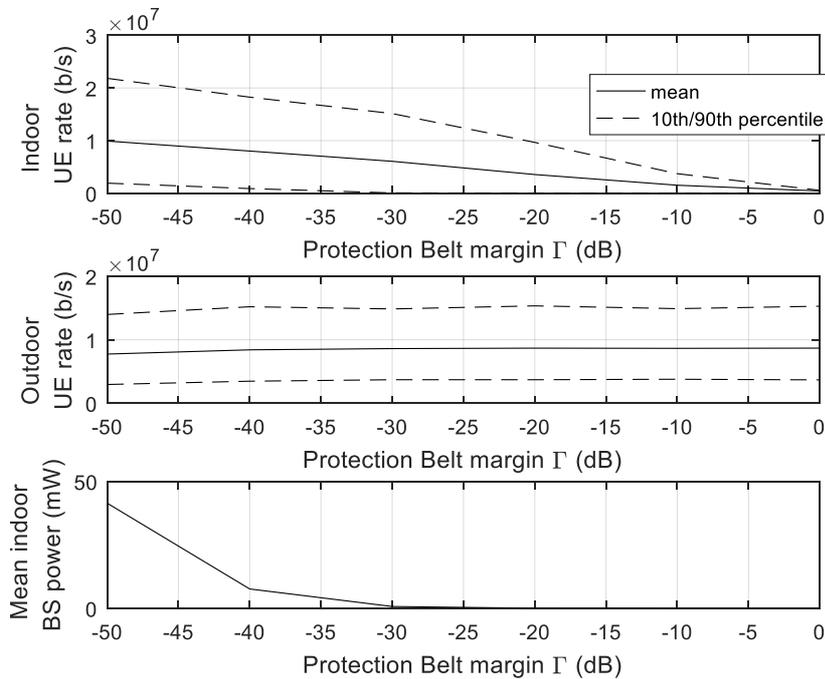

Figure 5. Indoor/outdoor UEs performance as a function of Protection Belt margin in the "Semi-static, REM-based protection" scenario.

Figure 5 presents indoor and outdoor UEs rates and mean indoor BS power for the "Semi-static, REM-based protection" scenario. Let us recall that for regulatory-based protection, i.e., modified LSA, the proper protection belt margin is -3 dB. It is visible that this will be much too restrictive. Indoor BSs will use nearly no power resulting in nearly no indoor transmission. The proper choice of the protection belt margin, providing a degradation of the outdoor UEs rate of 5-10%, is -40 dB. The $10^{th}$ percentile rate decreases by about 6% from 3.69 Mbps to 3.48 Mbps. The mean indoor BS power equals 7.7 mW. For comparison, the mean indoor BS power in the modified LSA scheme equals 0.0015 mW.

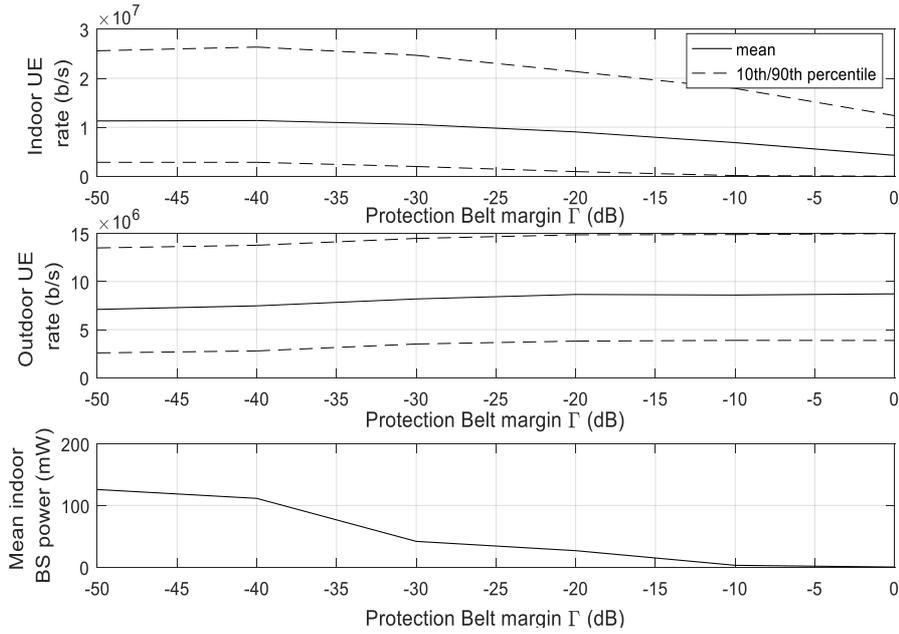

Figure 6. Indoor/outdoor UEs performance as a function of Protection Belt margin in the "Semi-static, REM-based protection with known protection area" scenario.

Similar results obtained for semi-static REM-based protection with known protection area are presented in Figure 6. In this case, higher mean indoor BS power is expected, as the system has knowledge on the possible location of outdoor UEs. In this case, the protection belt margin equals -30 dB. The $10^{th}$ percentile rate decreases from 3.87 Mbps to 3.5 Mbps, i.e., by about 9%. The mean indoor BS power equals 42 mW.

Most importantly, the two obtained Protection Belt Margins are valid for the considered environment, users distribution, etc., only.

Knowing the transmit power, location and pathloss model, a PAL protection area can be calculated for the CBRS scenario. The maximum received power out of both PAL transmitters is depicted on the map visible in Figure 7. As the PAL transmission is protected in each location where the wanted signal is higher than -93 dBm/ 20 MHz, only the indoor, "upper part" can be used by the GAA system. The only active indoor BS is the one with number 7 in Figure 4. The optimization of its power guaranteeing interference induced into the PAL protection area not higher than -77 dBm/ 20 MHz results in the mean indoor BSs power of 0.000088 mW. This is much less than in the "modified LSA" scenario. This is caused by lack of knowledge about the utilized indoor/outdoor network area available in the "modified LSA" scheme.

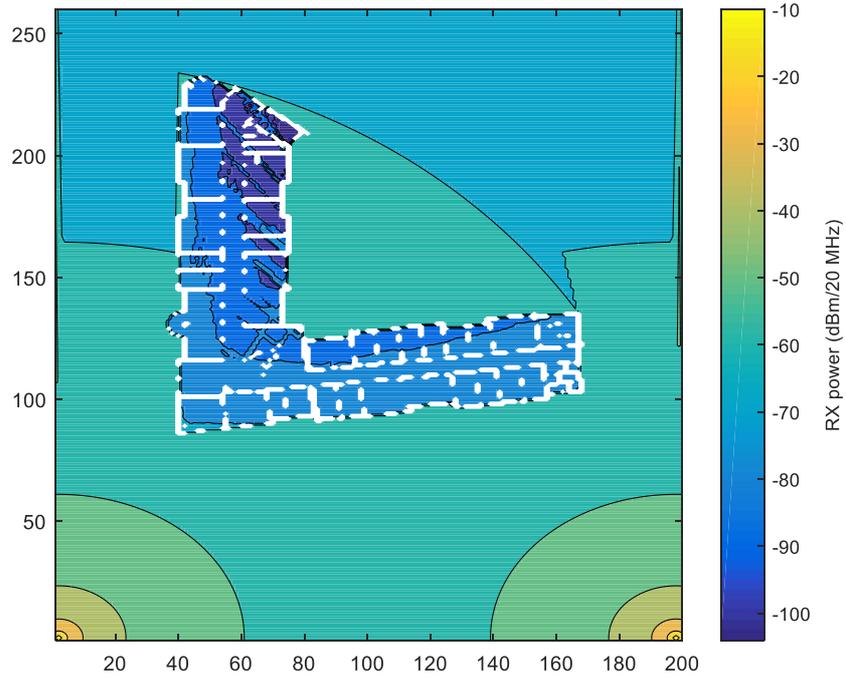

Figure 7. Received power from outdoor BSs.

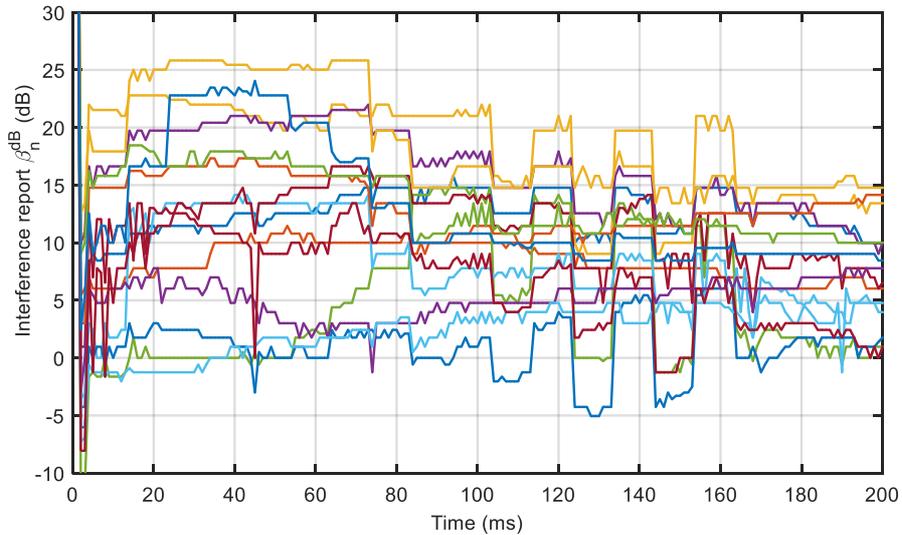

Figure 8. Illustration of interference reports from 15 outdoor UEs in the "dynamic REM" scenario.

In the case of the "dynamic REM" scenario, the important factors are how accurate and how often the interference reports are and how often the indoor BSs are reconfigured. It is assumed that the interference reports are calculated using $\Psi = 90$ in (7). An example of interference reports from 15 outdoor UEs sent every 1 ms is shown in Figure 8. It is assumed that REM introduces a 1 ms delay. Additionally, an update of the indoor BS power is run every 10 ms. At the beginning, all indoor BSs transmit with the maximum power, i.e., 21 dBm each. However, all outdoor UEs report a high possible increase of interference power, i.e., $\beta_n^{dB}$ much higher than 0 dB. This is caused by a delay in CQI reporting. At the beginning of the simulation there is no knowledge about the interference. However, this period is not considered while calculating the system statistics. After a few ms, many UEs report excessive interference power (negative $\beta_n^{dB}$ values). The first TX power adjustment is performed in 3$^{rd}$ ms. Many indoor BSs start to transmit with reduced power resulting in nearly no negative interference reports. The next BS power update is performed in the 13$^{th}$ ms (visible significant change in interference reports). As expected, many UEs report a possible significant increase of the received interference power. However, there is always at least one UE that is reporting interference power close to the threshold, i.e., 0 dB. This "worst case" user is protected by the proposed algorithm as well. Observe that the interference reports change in time as a result of pathloss/fading change.

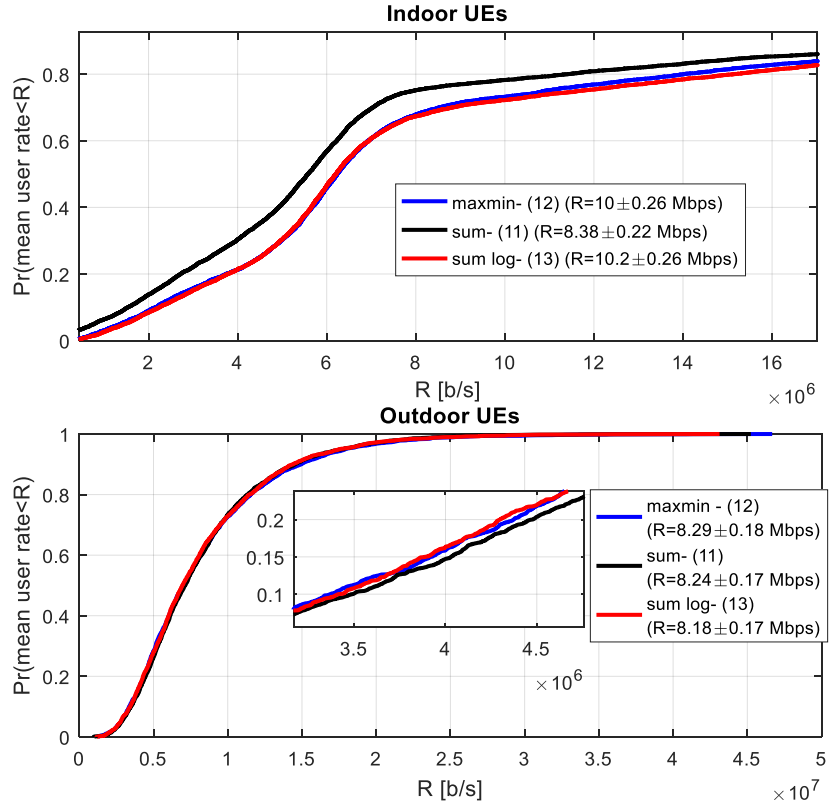

Figure 9. Comparison of indoor/outdoor UEs rate distribution for "dynamic REM" scenario using various goal functions.

In the case of the "dynamic REM" scenario it is important to find out the influence of the goal function $f(\ )$ defined in (10) on the achievable indoor/outdoor rate. Cumulative Density Functions of UE rates in this scenario using 3 goal functions defined in (11)-(13) are shown in Figure 9. Most importantly, the protection of outdoor UEs is not affected by the chosen goal functions. The 10$^{th}$ percentile rates overlap and the differences between the mean rates (shown in the legend) is within a 95% confidence interval (see numbers after ± in the legend). However, the indoor UE rates are significantly influenced by the choice of the goal function. The maximization of the sum of indoor TX power (black line) is the worst solution. The mean indoor UE rate can be increased by 1.82 Mbps if the sum of logarithms of TX power is maximized (red line). However, the optimal solution may change depending on the scenario, e.g., indoor UE distribution should be taken into account in order to provide them with the highest SINR and some load balancing among cells. Further results have been generated for the maximization of the summed indoor TX power, i.e., goal function (11), as a baseline approach in this scenario.

Knowing the optimal interference margin values, a comparison of all considered scenarios is possible. In Figure 10 the CDF of the observed mean rate of UE is shown. While semi-static REM-based protection is denoted by "REM sta.", its improvement with the knowledge of the protection area is denoted by "REM st.+". In the brackets, the mean users' rate is provided, together with a 95% confidence interval. Let us take a look at the degradation of the outdoor UE rate. It is visible that all REM-based schemes provide a similar level of outdoor UE protection, resulting in about 5% degradation of the $10^{th}$ percentile user mean rate in comparison to the indoor transmission being turned off. Modified LSA and CBRS schemes result in negligible or no degradation of the outdoor UE rate. On the other hand, the indoor UE rates in these scenarios are significantly lower than in the REM-based ones. It confirms the advantage of rich context information utilization using REM. As the regulatory-based CBRS scheme is significantly outperformed by the modified LSA scheme in terms of the indoor UE rate, this approach will not be considered in the next simulations.

In REM-based approaches, the indoor UEs achieve relatively high throughput, i.e., the mean is higher than 8 Mbps. Expectedly, the rate achieved in semi-static protection with a known protection area is higher than in the basic semi-static, REM-based protection. This is thanks to the knowledge about the possible location of outdoor UEs during indoor BSs power allocation.

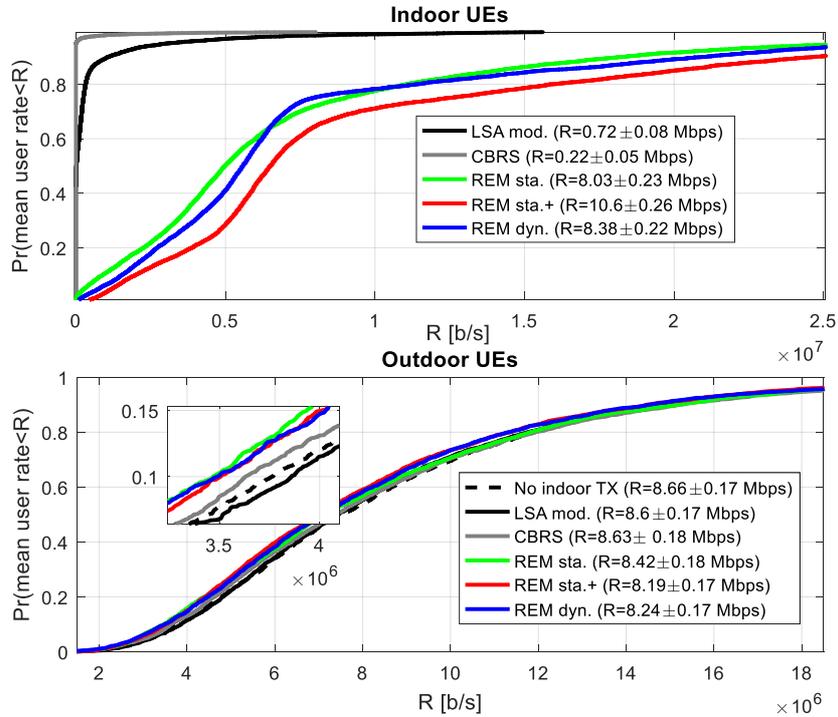

Figure 10. CDF of the mean rate of UEs

Most interestingly, the highest amount of context information available in dynamic REM-based protection is not reflected in the increased indoor UEs rate. First, it has been shown in Figure 9 that a correct choice of the goal function can improve the achievable indoor rate. Secondly, the dynamic protection of outdoor UEs is adjusted in time, while the semi-static scheme provides protection based on the mean rate of each outdoor UE. In the semi-static approach it is possible that at some time instant, the achievable outdoor UE rate will fall below the threshold of 90% of the rate with no indoor transmission.

In order to observe the above mentioned effect, the setup has been modified. Now, there is only one outdoor UE considered, moving along the building from left to right at 50 km/h. The total simulation time is 6 s, equivalent to 83 m of the UE path. Multipath effects are removed in order to allow the fast convergence of simulations. In each of 50 iterations, the only random variables are the locations of indoor UEs. The mean outdoor/indoor UE rate has been averaged over the period of 200 ms and all iterations. In semi-static protection schemes calibrations similar to the one presented in Figure 5 and Figure 6 have been performed, resulting in protection belt margins of -37.5 and -27.5 dB, respectively. The results are shown in Figure 11. The outdoor UE rate is most stable in dynamic REM-based protection and modified LSA. The no-REM solution sets relatively low indoor BSs power, which results in a relatively low mean indoor UE rate. In dynamic protection, the indoor BSs power is adjusted in time as a result of interference reports sent by the outdoor UE. It results in significant changes of the mean indoor UE rate. On the other hand, the mean indoor UE rate for semi-static protection is relatively stable over time. However, the outdoor UE rate changes significantly in time, achieving at some time instances a rate lower than 70% of the outdoor rate with no indoor transmission. This is a result of the outdoor UE being relatively close to one of the transmitting indoor BSs. Most importantly, the mean outdoor UE rate is similar in all REM-based solutions, constituting about 90%-95% of the rate achievable when there is no indoor transmission.

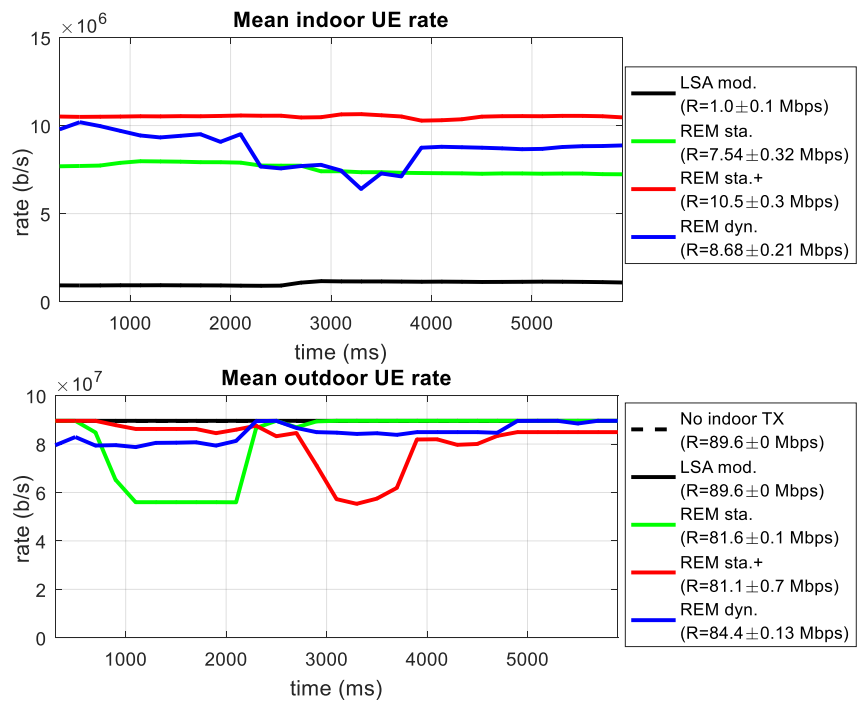

Figure 11. Mean indoor/outdoor UE rate in time in AWGN channel.

As the dynamic, REM-based scenario uses the highest amount of context information, it can be the most sensitive to some errors/inaccuracies, e.g., in the interference reports. First, in order to test the stability of the proposed approach, some delays are introduced to the baseline setup. The interference report from outdoor UE is delivered with a 1 ms delay to its BS. It is the same delay as introduced for CQI reports. However, the interference reports have to be sent to Operator A's REM and forwarded to Operator B's REM. This causes 1 second delay. As such, the algorithm calculating indoor BS power operates on the basis of outdated interference reports. The interference observed by a given UE at the time of indoor BS power update will be higher/lower than the one indicated in the interference report. The second degree of freedom is the period of indoor BS power update. Fast changes in interference reports should be reflected in fast indoor BSs reconfiguration. The influence of these delays on the UE rate is shown in Figure 12.

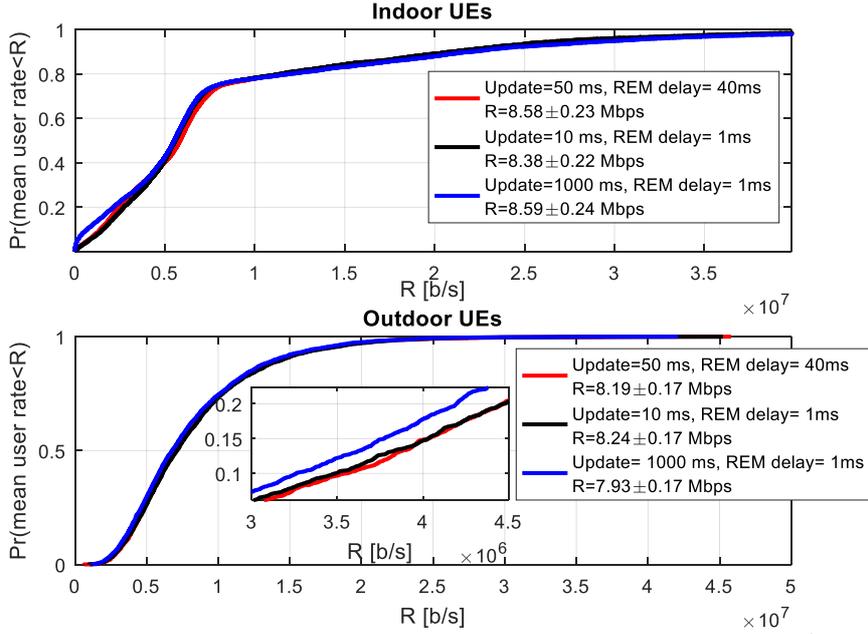

Figure 12. Influence of REM-based delay and updating delay on UE rate distribution in "dynamic, REM-based protection"

The most important issue is to provide continuous protection of outdoor UEs. While in the baseline system (REM delay of 1 ms and indoor BSs power update every 10 ms) the mean outdoor UE rate is 8.38 Mbps, it is nearly undegraded by a significant increase of REM delay (to 40 ms) and indoor BS reconfiguration period (to 50 ms). Probably the pedestrian outdoor UE channel is changing slowly enough to be followed by the indoor BS adaptation. For an extreme case (from the simulator perspective) of only one BS reconfiguration at the beginning of a 1000 ms period, the degradation of the outdoor UE rate is more significant. However, the mean outdoor UE rate is decreased by only about 4%.

The impact of outdoor UE location accuracy and quantization of interference reports in dynamic, REM-based protection is shown in Figure 13. A relative interference change has been reported using only 2 bits reflecting 4 possible interference power changes, i.e., {-6, -3, 3, 6} dB. Additionally, the outdoor UE locations have been determined with 50 m accuracy (red line). It is visible that these limitations do not significantly change the indoor/outdoor rate distribution. For a relatively stable environment (UEs being pedestrians) the recommended relative interference power change is within ±6 dB over the assumed indoor BS power update period of 10 ms. Therefore, 2 bits are enough for reporting the relative changes of interference. On the other hand, inaccuracy in UE location causes imperfection in the channel characterization between a given indoor BS and outdoor UE. However, the relative interference reporting and a 10 ms update period allow the algorithm to adapt to the environment. More detailed interference reports allow the algorithm to converge faster to the optimal solution.

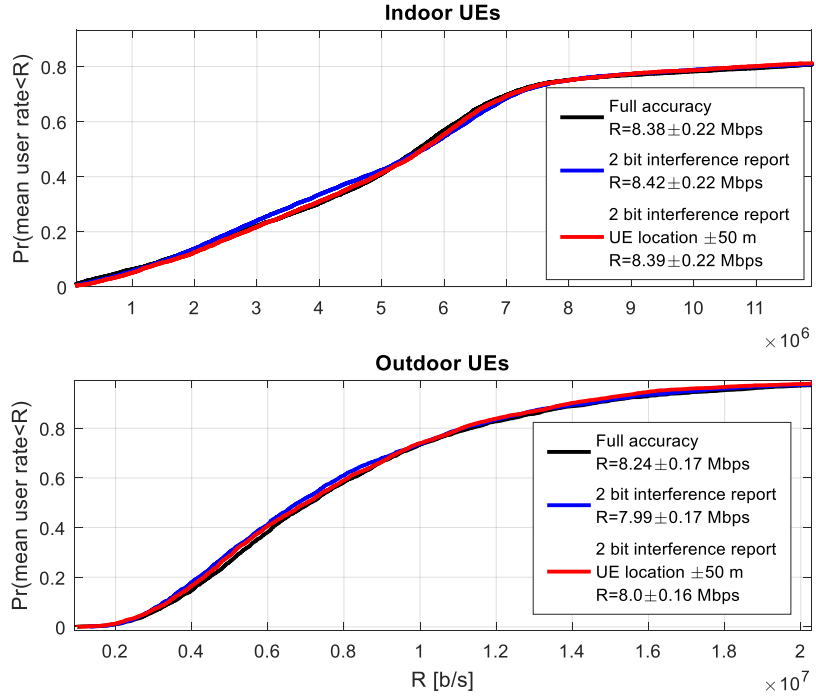

Figure 13. Influence of interference reports granularity on the indoor/outdoor UE rate distribution.

## Conclusions

In this paper REM-based schemes have been proposed for the allocation of base stations powers in indoor 4G/5G network, while considering interference generated to a licensed outdoor network. Results of a series of simulation studies of the tested indoor-outdoor environment show that utilization of context information via considered system of REMs enables effective dynamic spectrum access. The proposed solutions provide significant throughput increase for the indoor network as compared to the utilization of the two well-established regulatory-based approaches of modified LSA and CBRS. Moreover, the throughput gains of the proposed solutions are achieved while causing only limited interference to the outdoor mobile network. Thus, we conclude that REM-based dynamic spectrum access can be one of the key technologies that can be used to increase spectrum utilization in 5G networks.

## Conflicts of Interest

The authors declare that there is no conflict of interest regarding the publication of this paper.

## Funding Statement

This work was supported by the Polish Ministry of Science and Higher Education funds for the status activity project "Cognitive and sustainable communication systems", funds for the status activity for young scientists project DSMK/81/8136 and by Huawei Technologies, Sweden.

## Acknowledgments

The authors thank Magnus Isaksson and Per Tengkvist both with Huawei Technologies Sweden for their insightful discussions.
## References

[1] CISCO, "Cisco Visual Networking Index: Global Mobile Data Traffic Forecast Update, 2016-2021", white paper, available online: https://www.cisco.com/c/en/us/solutions/collateral/service-provider/visual-networking-index-vni/complete-white-paper-c11-481360.pdf (accessed 28.03.2018)

[2] J. Mitola, Cognitive Radio: An Integrated Agent Architecture for Software Defined Radio, Ph.D. thesis, KTH Royal Institute of Technology, 2000

[3] J. Mitola, Cognitive Radio Architecture Evolution, Proc. IEEE, vol. 97, no. 4, Apr. 2009

[4] A. Kliks, O. Holland, A. Basaure and M. Matinmikko, "Spectrum and license flexibility for 5G networks," in IEEE Communications Magazine, vol. 53, no. 7, pp. 42-49, July 2015, doi: 10.1109/MCOM.2015.7158264

[5] G. Ding et al., "Spectrum Inference in Cognitive Radio Networks: Algorithms and Applications," in IEEE Communications Surveys & Tutorials, vol. 20, no. 1, pp. 150-182, Firstquarter 2018, doi: 10.1109/COMST.2017.2751058

[6] W. Liang, S. X. Ng and L. Hanzo, "Cooperative Overlay Spectrum Access in Cognitive Radio Networks," in IEEE Communications Surveys & Tutorials, vol. 19, no. 3, pp. 1924-1944, thirdquarter 2017, doi: 10.1109/COMST.2017.2690866

[7] R. H. Tehrani, S. Vahid, D. Triantafyllopoulou, H. Lee and K. Moessner, "Licensed Spectrum Sharing Schemes for Mobile Operators: A Survey and Outlook," in IEEE Communications Surveys & Tutorials, vol. 18, no. 4, pp. 2591-2623, Fourthquarter 2016, doi: 10.1109/COMST.2016.2583499

[8] H.B. Yilmaz, T. Tugcu, F. Alagoz and S. Bayhan, "Radio environment map as enabler for practical cognitive radio networks," IEEE Communications Magazine, vol. 51, no. 12, pp. 162-169, December 2013

[9] Youping Zhao; Shiwen Mao; Neel, James O.; Reed, J.H., "Performance Evaluation of Cognitive Radios: Metrics, Utility Functions, and Methodology," in Proceedings of the IEEE , vol.97, no.4, pp.642-659, April 2009 doi: 10.1109/JPROC.2009.2013017

[10] J. van de Beek at al., "How a layered REM architecture brings cognition to today's mobile networks," in IEEE Wireless Communications, vol. 19, no. 4, pp. 17-24, August 2012 doi: 10.1109/MWC.2012.6272419

[11] Perez-Romero, J.; Zalonis, A.; Boukhatem, L.; Kliks, A.; Koutlia, K.; Dimitriou, N.; Kurda, R., "On the use of radio environment maps for interference management in heterogeneous networks," in IEEE Communications Magazine, vol.53, no.8, pp.184-191, August 2015

[12] D. Denkovski; V. Rakovic; M. Pavloski; K. Chomu; V. Atanasovski; L. Gavrilovska, "Integration of heterogeneous spectrum sensing devices towards accurate REM construction," in 2012 IEEE Wireless Communications and Networking Conference (WCNC), pp.798-802, 1-4 April 2012, doi: 10.1109/WCNC.2012.6214480
22

[13] Adrian Kliks, Leonardo Goratti, Tao Chen, "REM: Revisiting a Cognitive Tool for Virtualized 5G Networks", in 23rd International Conference on Telecommunications (ICT), 16-18 May 2016, Thessaloniki, Greece

[14] C. Liang, F. R. Yu and X. Zhang, "Information-centric network function virtualization over 5g mobile wireless networks", IEEE Network, vol. 29, no. 3, pp. 68-74, May-June 2015, doi: 10.1109/MNET.2015.7113228

[15] C. Liang and F. R. Yu, "Wireless Network Virtualization: A Survey, Some Research Issues and Challenges," in IEEE Communications Surveys & Tutorials, vol. 17, no. 1, pp. 358-380, First Quarter 2015, doi: 10.1109/COMST.2014.2352118

[16] A. Chowdhery; R. Chandra; P. Garnett; P. Mitchell, "Characterizing spectrum goodness for dynamic spectrum access," in 2012 50th Annual Allerton Conference on Communication, Control, and Computing (Allerton), pp.1360-1367, 1-5 Oct. 2012, doi: 10.1109/Allerton.2012.6483376

[17] Oliver Holland, Bernd Bochow; Konstantinos Katzis, "IEEE 1900.6b: Sensing support for spectrum databases," in 2015 IEEE Conference on Standards for Communications and Networking (CSCN), pp.199-205, 28-30 Oct. 2015

[18] M.M. Kassem; M.K. Marina, "Future wireless spectrum below 6 GHz: A UK perspective," in 2015 IEEE International Symposium on Dynamic Spectrum Access Networks (DySPAN), pp. 59-70, Sept. 29 2015-Oct. 2 2015, doi: 10.1109/DySPAN.2015.7343850

[19] Yun Li, "Grass-root based Spectrum Map database for self-organized cognitive radio and heterogeneous networks: Spectrum measurement, data visualization, and user participating model," in 2015 IEEE Wireless Communications and Networking Conference (WCNC), pp.117-122, 9-12 March 2015, doi: 10.1109/WCNC.2015.7127455

[20] Nan Wang; Yue Gao; Evans, B., "Database-augmented spectrum sensing algorithm for cognitive radio," in 2015 IEEE International Conference on Communications (ICC), pp. 7468-7473, 8-12 June 2015, doi: 10.1109/ICC.2015.7249520

[21] K. Sato and T. Fujii, "Kriging-Based Interference Power Constraint: Integrated Design of the Radio Environment Map and Transmission Power," in IEEE Transactions on Cognitive Communications and Networking, vol. 3, no. 1, pp. 13-25, March 2017, doi: 10.1109/TCCN.2017.2653189

[22] Z. Hou, Y. Zhou, L. Tian, J. Shi, Y. Li and B. Vucetic, "Radio Environment Map-Aided Doppler Shift Estimation in LTE Railway," in IEEE Transactions on Vehicular Technology, vol. 66, no. 5, pp. 4462-4467, May 2017, doi: 10.1109/TVT.2016.2599558

[23] S. Sodagari, "A Secure Radio Environment Map Database to Share Spectrum," in IEEE Journal of Selected Topics in Signal Processing, vol. 9, no. 7, pp. 1298-1305, Oct. 2015, doi: 10.1109/JSTSP.2015.2426132

[24] P. Tengkvist, G. P. Koudouridis, C. Qvarfordt, M. Dryjanski and M. Cellier, "Multi-dimensional radio service maps for position-based self-organized networks," 2017 IEEE 22nd International Workshop on Computer Aided Modeling and Design of Communication Links and Networks (CAMAD), Lund, 2017, pp. 1-6., doi: 10.1109/CAMAD.2017.8031530

[25] K. Katagiri, K. Sato, T. Fujii, "Crowdsourcing-Assisted Radio Environment Database for V2V Communication." Sensors 18, no. 4: 1183, 2018

[26] M. Sybis, P. Kryszkiewicz, P. Sroka "On the Context-Aware, Dynamic Spectrum Access for Robust Intraplatoon Communications" in Mobile Information Systems, vol. 2018, Article ID 3483298, doi: 10.1155/2018/3483298

[27] Requirements for Commercial Operation in the U.S. 3550-3700 MHz Citizens Broadband Radio Service Band, Working Document WINNF-TS-0112, Wireless Innovation Forum, 1 May 2018.

[28] CEPT "Technical sharing solutions for the shared use of the 2300-2400 MHz band for WBB and PMSE" CEPT Report 58, July 2015.

[29] N. Morozs, T. Clarke and D. Grace, "Heuristically Accelerated Reinforcement Learning for Dynamic Secondary Spectrum Sharing," in *IEEE Access*, vol. 3, pp. 2771-2783, 2015. doi: 10.1109/ACCESS.2015.2507158

[30] G. P. Koudouridis and P. Soldati, "Spectrum and Network Density Management in 5G Ultra-Dense Networks," in IEEE Wireless Communications, vol. 24, no. 5, pp. 30-37, October 2017, doi: 10.1109/MWC.2017.1700087

[31] G. Caso, L. D. Nardis and M. G. D. Benedetto, "Toward Context-Aware Dynamic Spectrum Management for 5G," in IEEE Wireless Communications, vol. 24, no. 5, pp. 38-43, October 2017, doi: 10.1109/MWC.2017.1700090

[32] D. B. Rawat, C. Bajracharya and S. Grant, "nROAR: Near Real-Time Opportunistic Spectrum Access and Management in Cloud-Based Database-Driven Cognitive Radio Networks," in IEEE Transactions on Network and Service Management, vol. 14, no. 3, pp. 745-755, Sept. 2017, doi: 10.1109/TNSM.2017.2730201

[33] ICT-317669-METIS project, Deliverable D5.4, "Future spectrum system concept", deliverable of the FP7 project Mobile and wireless communications Enablers for Twenty-twenty (2020) Information Society (METIS), April 2015 available at: https://www.metis2020.com/wp-content/uploads/deliverables/METIS_D5.4_v1.pdf

[34] METIS-II project, Deliverable D3.2, "Enablers to secure sufficient access to adequate spectrum for 5G", deliverable of the Mobile and wireless communications Enablers for Twenty-twenty (2020) Information Society – II (METIS-II) project, under grant agreement 671680, June 2017, available at https://metis-ii.5g-ppp.eu/wp-content/uploads/deliverables/METIS-II_D3.2_V1.0.pdf

[35] ICT-671639 COHERENT, Deliverable D4.2, "Final report on flexible spectrum management", December 2017, available at: http://www.ict-coherent.eu/coherent/wp-content/uploads/2018/03/COHERENT_D4.2_v8.pdf

[36] M. Mustonen, M. Matinmikko, D. Roberson and S. Yrjola, ``Evaluation of recent spectrum sharing models from the regulatory point of view,'' 1st International Conference on 5G for Ubiquitous Connectivity, Akaslompolo, 2014, pp. 11-16

[37] Reconfigurable Radio Systems (RRS); Information elements and protocols for the interface between LSA Controller (LC) and LSA Repository (LR) for operation of Licensed Shared Access (LSA) in the 2300 MHz-2400 MHz band, ETSI TS 103 379, V0.0.5 (2016-03).

[38] R. H. Tehrani, S. Vahid, D. Triantafyllopoulou, H. Lee and K. Moessner, "Licensed Spectrum Sharing Schemes for Mobile Operators: A Survey and Outlook," in IEEE Communications Surveys & Tutorials, vol. 18, no. 4, pp. 2591-2623, Fourthquarter 2016, doi: 10.1109/COMST.2016.2583499

[39] C. Galiotto, G. K. Papageorgiou, K. Voulgaris, M. M. Butt, N. Marchetti and C. B. Papadias, "Unlocking the Deployment of Spectrum Sharing With a Policy Enforcement Framework," in IEEE Access, vol. 6, pp. 11793-11803, 2018, doi: 10.1109/ACCESS.2018.2799244

[40] Imtiaz Parvez, M. G. S. Sriyananda, İsmail Güvenç, Mehdi Bennis, and Arif Sarwat, "CBRS Spectrum Sharing between LTE-U and WiFi: A Multiarmed Bandit

Approach," Mobile Information Systems, vol. 2016, Article ID 5909801, 12 pages, 2016. doi:10.1155/2016/5909801

[41] M. M. Sohul, M. Yao, T. Yang and J. H. Reed, "Spectrum access system for the citizen broadband radio service," in IEEE Communications Magazine, vol. 53, no. 7, pp. 18-25, July 2015, doi: 10.1109/MCOM.2015.7158261

[42] A. Kliks, P. Kryszkiewicz, Ł. Kulacz, K. Kowalik, M. Kołodziejski, H. Kokkinen, J. Ojaniemi, A. Kivinen, "Application of the CBRS model for wireless systems coexistence in 3.6-3.8 GHz band", in proc. of 12th International Conference Cognitive Radio Oriented Wireless Networks, CROWNCOM 2017, Lisbon, Portugal, September 20–21, 2017

[43] Adrian Kliks, Andreas Zalonis, Ioannis Dagres, Andreas Polydoros, and Hanna Bogucka "PHY Abstraction Methods for OFDM and NOFDM Systems" Journal of Telecommunications and Information Technology 2009

[44] Michael Grant and Stephen Boyd. CVX: Matlab software for disciplined convex programming, version 2.0 beta. http://cvxr.com/cvx, September 2013.

[45] Zakaria Hanzaz, Hans Dieter Schotten „Performance Evaluation of Link to System Interface for Long Term Evolution System" International Wireless Communications and Mobile Computing Conference (IWCMC), 2011.

[46] Bartosz Bossy, Paweł Kryszkiewicz, Hanna Bogucka „Optimization of Energy Efficiencyin the Downlink LTE Transmission" IEEE International Conference on Communications 2017

[47] Mohamad Yassin, Mohamed A. AboulHassan, Samer Lahoud, Marc Ibrahim, Dany Mezher, Bernard Cousin, Essam A. Sourour, "Survey of ICIC techniques in LTE networks under various mobile environment parameters", Wireless Networks, February 2017, Volume 23, Issue 2, pp 403–418

[48] R. Kwan, C. Leung, and J. Zhang, "Proportional fair multiuser scheduling in LTE," IEEE Signal Processing Letters, vol. 16, no. 6, pp. 461–464, June 2009.

[49] IST-4-027756 WINNER II, Deliverable D1.1.2 V1.2, "WINNER II Channel Models", September 2007, available at: https://www.cept.org/files/8339/winner2%20-%20final%20report.pdf